\documentclass[twocolumn]{aastex631}

\shorttitle{The luminosity-area relation of quasars' nebulae}
\shortauthors{Arrigoni Battaia et al.}
\graphicspath{{./}{figures/}}

\begin{document}

\title{The luminosity-area relation of $z>2$ quasars' Ly$\alpha$ nebulae}

\author[0000-0002-4770-6137]{Fabrizio Arrigoni Battaia}
\affiliation{Max-Planck-Institut f\"ur Astrophysik, Karl-Schwarzschild-Str. 1, D-85748 Garching bei M\"unchen, Germany}

\author[0000-0003-4196-8555]{Aura Obreja}
\affiliation{Universit\"ats-Sternwarte M\"unchen, Scheinerstraße 1, D-81679 M\"unchen, Germany}

\author[0000-0002-6748-2900]{Tiago Costa}
\affiliation{Max-Planck-Institut f\"ur Astrophysik, Karl-Schwarzschild-Str. 1, D-85748 Garching bei M\"unchen, Germany}

\author[0000-0002-6822-2254]{Emanuele P. Farina}
\affiliation{Gemini Observatory, NSF's NOIRLab, 670 N A'ohoku Place, Hilo, Hawai'i 96720, USA}

\author[0000-0001-8467-6478]{Zheng Cai}
\affiliation{Department of Astronomy, Tsinghua University, Beijing 100084, China}



\begin{abstract}

Cool ($T\sim10^4$~K) gas is commonly observed around $z>2$ quasars as traced by extended Ly$\alpha$ emission. These large-scale nebulae are usually studied using circularly averaged surface brightness profiles, which  
suppress information on morphological differences. 
Here, we revisit the Ly$\alpha$ nebulae around 78 $z\sim2-3$ quasars to obtain a novel estimate of their area and asymmetry using a common redshift-corrected surface-brightness threshold.
We find a luminosity-area relation of the form ${{\rm log}(L_{\rm Ly\alpha}^{\rm Neb})=a_1 log({\rm Area^{Neb})+a_0}}$. 
Most nebulae are symmetric and bright, the most lopsided ones being the faintest and the less extended. The Enormous Lyman-Alpha Nebulae, asymmetric due to the presence of active companions, are the exceptions to this trend. 
By using simulations able to reproduce $z\sim6$ quasar's nebulae, we show that the observed relation should not vary with redshift.  
Finally, we discuss possible mechanisms that drive the relation and future work needed to constrain them.

\end{abstract}



\section{Introduction} \label{sec:intro}

Quasars are the brightest episodes of active galactic nuclei (AGN), with 
bolometric luminosities up to $L_{\rm bol}\sim10^{50}$~erg~s$^{-1}$ (e.g., \citealt{Shen2020}). Their activity is sustained by large reservoirs of multiphase gas, funneling from
large-scales to the host galaxies, and ultimately to the accretion disks of central black holes.

The surrounding cool gas at $T\sim10^4$~K has been long ago predicted to shine as a Ly$\alpha$ glow around high-redshift ($z$) quasars (QSO) (e.g., \citealt{Rees1988}). Pioneering studies were able to uncover  
this signal (e.g., \citealt{HuCowie1987}), study the gas physical properties (e.g., \citealt{heckman91a}), infer inflowing motions (e.g., \citealt{Weidinger05}), and investigate correlations between nebulae and QSO properties (e.g., \citealt{Christensen2006}). Notwithstanding these results, those works were hampered by the inherent limitations of those observations, e.g., small field of views, limited sensitivities (e.g., \citealt{Christensen2006}).

A trasformational era in the detection of such large-scale gas reservoirs is ongoing, thanks to  
technical developments especially at optical wavelengths, where new sensitive integral field units (IFUs) operate. An ever increasing number of works (e.g., \citealt{Borisova2016,
Cai2018,OSullivan2020,Fossati2021,Lau2022}) use the Multi-Unit Spectroscopic Explorer (MUSE; \citealt{Bacon2010}) and the Keck Cosmic Web Imager (KCWI; \citealt{Morrissey2012}) to routinely study 
halo gas even around some of the highest $z$ objects ($z>6$; e.g., \citealt{Farina2019}).  
These studies find Ly$\alpha$ nebulae with very diverse morphologies extending, in the most extreme cases, out to hundreds of kiloparsecs. Most of these nebulae show rather quiescent kinematics (FWHM$\sim600$~km~s$^{-1}$; \citealt{FAB2019}, hereafter FAB19; see e.g., \citealt{Borisova2016,Ginolfi2018,Travascio2020} for some exceptions) consistent with gravitational motions inside the expected massive QSOs' dark-matter (DM) halos  ($M_{\rm DM}\sim10^{12.5}$~M$_{\odot}$; e.g., \citealt{white12,Timlin2018}).  
The bulk of this Ly$\alpha$ emission therefore traces cool dense gas inspiraling into the QSO's halo (e.g., \citealt{FAB2018}).

The extreme luminosity of QSOs makes it very hard to observe their much fainter host galaxies. However, the properties of the
latter (distribution of atomic and molecular gas, and  
galaxies inclination with respect to the line of sight) determine to a great extent the final observed appearance of the nebulae (e.g., \citealt{Costa2022}). 
Other important factors that define the observed glows are: the configuration of the cool gas within the host DM halos, the geometry of the QSO's ionization cones, and Ly$\alpha$ resonant scattering. 
Therefore, any clear relation between the physical properties of these nebulae can greatly help in understanding their nature. 

Specifically, despite the very diverse nebulae morphology, circularly-averaged Ly$\alpha$ surface brightness (SB) profiles are self-similar at all surveyed redshifts \citep[$2<z<6.7$, FAB19; ][]{Farina2019,Cai2019,Fossati2021}. These SB profiles suppress information on the morphological differences of the nebulae, which could be crucial to understand their nature. 
Recent simulations (\citealt{Costa2022}), able to reproduce the Ly$\alpha$ emission around $z\sim6$ QSOs, show that the  
QSO host-galaxy inclination with respect to our line-of-sight  
could determine how we see the extended Ly$\alpha$ glow as QSO-powered outflows create a path of least resistance for Ly$\alpha$ photons along the host-galaxy minor axis. The expectation is to see fainter and more lopsided nebulae when the galaxy is close to edge-on, and brighter and more symmetric nebulae when the galaxy is close to face-on, possibly driving a luminosity–area relation for the Ly$\alpha$ nebulae. 
Importantly, this prediction is in place when taking into account the resonant scattering of Ly$\alpha$ photons.

Here, we revisit $z\sim2-3$ observations (\citealt{cantalupo14,hennawi+15}; FAB2019; \citealt{Cai2019}) targeting type-I QSOs (i.e. those having their ionization cone directed towards us, in accord with AGN unified models; e.g. \citealt{Anton93})
to show that Ly$\alpha$ nebulae populate a very tight power-law relation between their luminosity and area, given a common, dimming-corrected SB. This relation, together with a novel estimate of asymmetry, capture the morphological differences of nebulae more reliably and therefore enable closer comparison with simulations.

This paper is organized as follows. Section~\ref{sec:observations} 
describes the observational sample. 
We select a common dimming-corrected SB threshold for all QSO nebulae,  
and obtain luminosities, areas, and asymmetries. 
Section~\ref{sec:theRelation} presents the 
luminosity-area relation and its fit. 
Section~\ref{sec:comparison} shows that the relation holds out to $z\sim6$ by using the simulations of \citet{Costa2022}. Then, we briefly discuss  
possible mechanisms driving the relation and  
observations to constrain them (Section~\ref{sec:drivers}).
Finally, Section~\ref{sec:summary} summarizes our work.  
We adopt the cosmological parameters $H_0 = 70$~km~s$^{-1}$~Mpc$^{-1}$, $\Omega_{\rm M} =0.3$, and $\Omega_{\Lambda} =0.7$. In this cosmology, $1\arcsec$ is 8.4 (7.4) physical kpc at $z=2.0412$ (3.406)
(lowest and highest redshift for the observations).

\section{Observational Sample}
\label{sec:observations}

We rely on the $z\sim2-3$ type-I QSOs' nebulae observed with MUSE and KCWI by FAB19 (61 objects at $3.0<z<3.5$, including 1 enormous Ly$\alpha$ nebula (ELAN); \citealt{FAB2018}), \citet{Cai2019} (16 objects at $2.1<z<2.3$), respectively, and two $z\sim2$ ELANe observed with narrow-band imaging (\citealt{cantalupo14,hennawi+15}), representing all the systems for which we have data at the moment. For each system, we use the Ly$\alpha$ SB maps obtained in the original works. We rely on the optimally extracted maps for the IFU data and on maps obtained with $\approx30$\AA\ filters for the other works. While these systems have been observed with different instruments and down to different sensitivities, we use a conservative estimate of their area and luminosity, as explained in the next section. Importantly, the unresolved QSO emission has been subtracted before generating the SB maps for the IFU data (FAB19, \citealt{Cai2019}), while for the NB data this process cannot be reliably performed (\citealt{cantalupo14,hennawi+15}).
The targeted QSOs 
have 1450~\AA\ rest-frame magnitudes in the range $-25.3 \leq M_{\rm1450} \leq -28.9$, which translate to bolometric luminosities $10^{46.7} \leq L_{\rm bol} \leq 10^{48}\ {\rm erg\ s^{-1}}$ (\citealt{Runnoe2012}). All QSOs' properties are given as online material. Testing and expanding our results with larger samples of QSOs will be the scope of future works.

\subsection{Selection of a common surface-brightness threshold}
\label{sec:cut}

In an expanding universe, the SB of similar objects 
decreases very rapidly with redshift as $(1+z)^{-4}$ (\citealt{Tolman1930,Tolman1934}).  
Therefore, similarly deep observations, for example those targeting $z\sim3$ and $z\sim6$ QSOs' nebulae by FAB19 and \citet{Farina2019}, will be sensitive to very different SB levels.  
Thus, to compare nebulae areas and luminosities at different redshifts, we  use a common dimming-corrected SB threshold.  

\begin{figure*}
    \centering
    \includegraphics[width=1.0\textwidth]
    {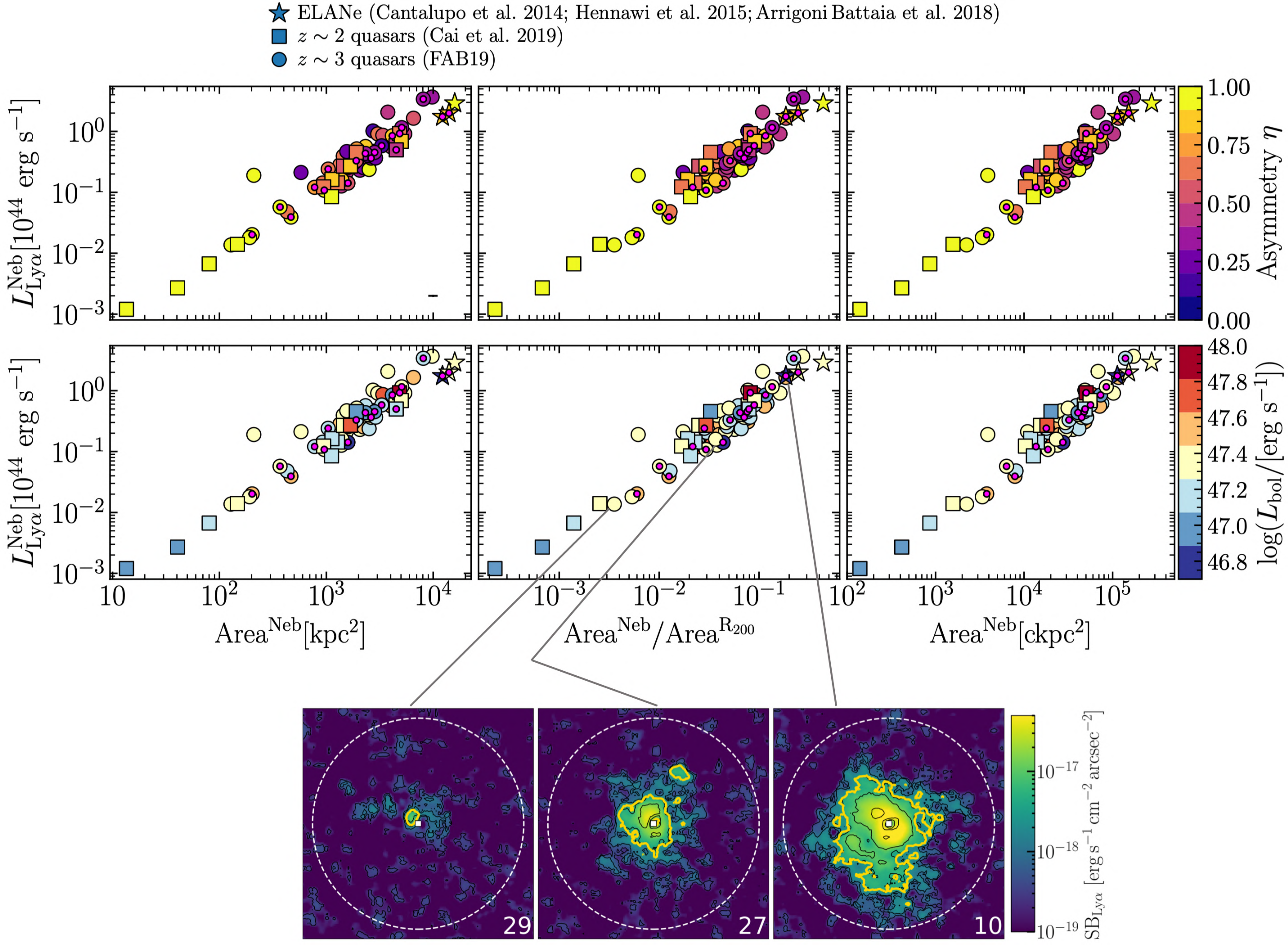}
    \caption{\textit{Top:} Luminosity vs physical (left), normalized (center), and comoving (right) area 
    of the QSOs' Ly$\alpha$ nebulae.  
    The data-points are color-coded following the asymmetry $\eta$. 
    Small magenta dots indicate  
    radio-loud quasars. The errorbar in the bottom right corner of the left plot indicates the median error of about 10\% and 4\% on area and luminosity, respectively. \textit{Center:} Same as the top panels, but with data-points color-coded following the QSOs
    bolometric luminosity $L_{\rm bol}$. 
     \textit{Bottom:} Ly$\alpha$ SB maps (FAB19)
     of three nebulae with 
     QSOs having $L_{\rm bol}\sim10^{47.5}$~erg~s$^{-1}$. The 
     IDs from QSO MUSEUM are indicated in the lower-right corner. The reference isophote  
     for the current analysis is highlighted in yellow. The white circle denotes the  
     virial radius for a DM halo of $M_{\rm DM}=10^{12.5}$~M$_{\odot}$ at 
     $z=3$ ($R_{\rm vir}\sim107$~kpc).}
    \label{Fig1}
\end{figure*}

We select a SB threshold that allows us to have all Ly$\alpha$ nebulae detected for the QSO MUSEUM survey and all but one (Q0848-0114) for the survey by \citet{Cai2019}. This reference SB isophote is an observed SB$_{\rm Ly\alpha}=1.23\times10^{-17}$~erg~s$^{-1}$~cm$^{-1}$~arcsec$^{-2}$ at the redshift of the Jackpot ELAN ($z=2.0412$; \citealt{hennawi+15}), the lowest $z$ in our sample, 
and corresponds to $\sim3.5\times10^{-18}$~erg~s$^{-1}$~cm$^{-1}$~arcsec$^{-2}$ at the median $z$ for QSO MUSEUM ($z=3.17$). We exclude the $z\sim6$ sample by \citet{Farina2019}  
as their observations do not achieve the sensitivity needed to detect signal at levels corresponding to the selected SB threshold (i.e., $3.9\times10^{-19}$~erg~s$^{-1}$~cm$^{-1}$~arcsec$^{-2}$ at $z=6.2$).
A higher SB reduces the statistics at lower $z$. Importantly, we tested two additional SB thresholds for $2\times$ and $4\times$ higher values, resulting in samples of 75 and 66 objects, respectively.

We then estimate  
the area and the luminosity of each QSO's 
Ly$\alpha$ nebula within the isophote corresponding to the chosen dimming-corrected SB threshold. In our analysis we 
adopt three types of area: physical, physical but normalized by the projected area of a $10^{12.5}$~M$_{\odot}$ DM halo at the redshift of the source, and comoving. The latter two areas allow for a better comparison of nebulae at different redshifts, taking into account the cosmological growth of structures, as done e.g., in the analysis of their SB profiles. Regarding the halo mass 
needed for normalization, we assume a typical value found from clustering studies (e.g., \citealt{white12,Timlin2018}) and also confirmed by considerations on the line-width of the extended Ly$\alpha$ emission (FAB19; \citealt{Farina2019}). For reference, a $10^{12.5}$~M$_{\odot}$ DM halo at $z=3$ have a virial radius $R_{\rm vir}=107$~kpc, corresponding to a projected area of $\approx36000$~kpc$^2$.

In our analysis we exclude the region of 1 arcsec$^2$ used for the normalization of the QSO unresolved emission for IFU data, and the region (two times the seeing) contaminated by unresolved QSO emission in the two ELANe with narrow-band data. 

The redshifts of the targeted QSOs are uncertain.  
We used the redshift from the extended Ly$\alpha$ nebula, except for 
two systems  
(\citealt{cantalupo14,hennawi+15}). 
However, 
the uncertainties on the redshifts translates to very minor differences in the SB threshold for each object, with a maximum variation of $\sim8$\%, and a mean of $\sim2$\%.
The uncertainties on the areas depend on these redshift errors and on the specific value of individual pixels, and are estimated to be negligible ($\sim10\%$). Any systematic variation for the area (e.g., due to flux calibration) affects also the luminosities and therefore only mildly impacts our results.

With the reference SB cut, the nebulae have Ly$\alpha$ luminosities of 
$1.2\times10^{41}\leq L_{\rm Ly\alpha}^{\rm Neb}\leq 3.67\times10^{44}$~erg~s$^{-1}$ and physical areas from $13$ to $15902$~kpc$^2$, corresponding to comoving areas $140\leq {\rm Area^{\rm Neb}} \leq 275721$~ckpc$^2$. The three ELANe, known to host multiple AGN and dusty star-forming galaxies (\citealt{cantalupo14,hennawi+15,FAB2018,FAB2021,TC2021}), are the only systems with physical areas exceeding $10^4$~kpc$^2$, while in comoving areas they are within the top 8\% of the sample.

The nebulae have different degrees of asymmetry with respect to the QSO position. We quantify this asymmetry by finding the direction (passing through the QSO position) that maximizes the area of nebula emission on one side. The asymmetry is then parametrized by the ratio $\eta = (A^{\rm neb}_{\rm max} - A^{\rm neb}_{\rm min})/A^{\rm neb}_{\rm max}$, where $A^{\rm neb}_{\rm max}$ and $A^{\rm neb}_{\rm min}$ are the areas on the maximum and minimum side. This ratio ranges between 0 and 1, corresponding to the symmetric and lopsided cases, respectively. The asymmetry value depends on the used SB threshold, with the smaller nebulae at each SB cut being usually the more lopsided (see Section~\ref{sec:theRelation}). We provide all the properties of the observed Ly$\alpha$ nebulae as online material.

\section{The luminosity-area relation of high-redshift quasar's nebulae}
\label{sec:theRelation}

Figure~\ref{Fig1} shows luminosity versus area for these QSOs' Ly$\alpha$ nebulae using 
the three different area definitions.
The data covers 3.5 orders of magnitude in both 
luminosity and area, but most observations ($\sim70$~\%) 
are clustered in the region $10^{43}\,{\rm erg\, s^{-1}\,}\lesssim L_{\rm Ly\alpha}^{\rm Neb}\lesssim 10^{44}$~erg~s$^{-1}$, physical areas $7\times10^2\,{\rm kpc^{2}\,}\lesssim \rm Area^{\rm Neb}\lesssim 7\times10^3$~kpc$^{2}$, normalized areas $0.02\lesssim \rm Area^{\rm Neb}\lesssim 0.2$, and comoving areas ${10^4\,{\rm ckpc^{2}\,}\lesssim \rm Area^{\rm Neb}}\lesssim 10^5$~ckpc$^{2}$.
Overall, all the datapoints are on a well-defined luminosity-area relation, which does not depend on $L_{\rm bol}$ for the current sample. 

The relation is instead driven by more lopsided nebulae at smaller areas and lower luminosities (top panel). An example of nebulae SB maps along this trend for fixed $L_{\rm bol}\sim10^{47.5}$~erg~s$^{-1}$ is shown in the bottom panel of Figure~\ref{Fig1}. An exception to this trend are the ELANe, which have some of the largest areas but 
$\eta\sim1$, likely emphasizing the importance of active companions in powering the emission in these objects (e.g., \citealt{FAB2021}). 

\begin{figure}
    \centering
    \includegraphics[width=0.46\textwidth]{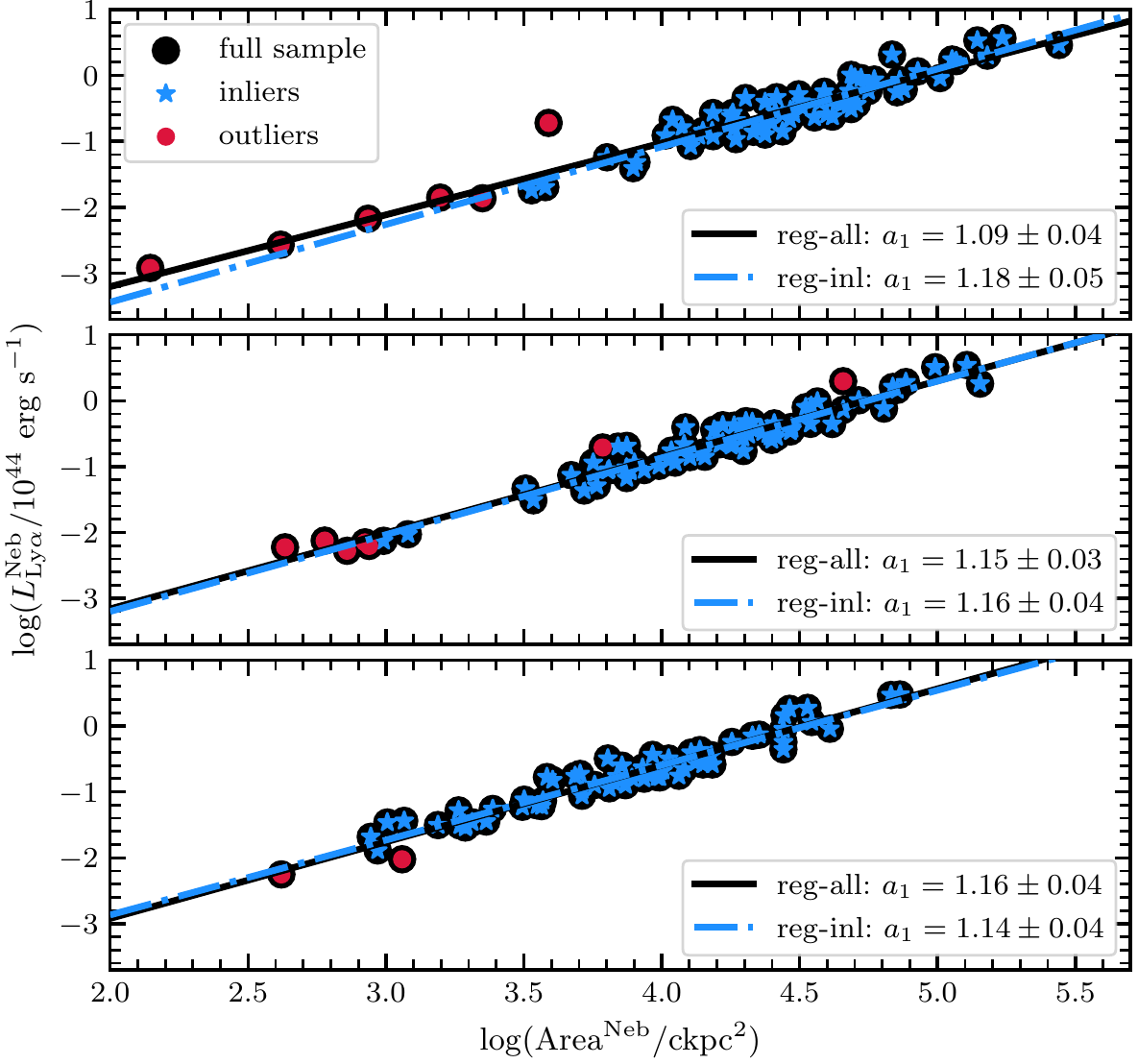}
    \caption{Linear regressions for the case with comoving areas. The linear regression through all the observations (solid black line, black circles) and through the observations classified as inliers by MCD (dashed-dotted blue line,  blue stars) for the reference SB threshold (top), and for a $2\times$ (middle) and $4\times$ (bottom) higher SB threshold. The red circles are the MCD outliers.
    The legends  
    list the obtained slopes.} 
    \label{Fig2}
\end{figure}

We find the linear regression for each luminosity-area relation and for the aforementioned three SB cuts. We report all the parameters of the relations in Table 1. Here we discuss our procedure in more detail for the comoving areas. The other cases are very similar.

Figure~\ref{Fig2} (top) shows again the relation with comoving areas as in Figure~\ref{Fig1}, but in this case split in outliers 
and inliers, 
as classified by the Minimum Covariance Determinant (MCD) estimator \citep{Rousseeuw1984,Rousseeuw1985}. This estimator is known to accurately identify outliers in multi-variate normally-distributed data. The linear regression through the complete dataset is consistent within $2\sigma$
to the one through inliers only. 
The correlation coefficients of the two regressions are both very large (0.96 and 0.93, respectively). 
Therefore, the regression parameters for the whole sample are influenced by the few data points at low luminosities and low areas, which could be slightly biased high due to instrumental uncertainties (e.g., pixel size).
The middle and bottom panels of Figure~\ref{Fig2} show that the use of a $2\times$ and $4\times$ higher  SB threshold results, as expected, in an increase of the normalisation, 
but the slope remains consistent 
to the value for the reference SB threshold. 
We note that the linear regressions for comoving  
and normalized areas have the exact same slopes. This is due
to the fact that for a fixed halo mass, the virial radius is roughly $\propto (1+z)^{-1}$.

Uncertainties in the quasar's redshifts do not affect the luminosity-area relation  
because its slope 
is close to 1, and hence uncertainties in redshift move Ly$\alpha$ nebulae along the relation. 

Finally, we stress that the median $L_{\rm Ly\alpha}^{\rm Neb}$ for $z\sim3$ nebulae is $\sim2.5\times$ that of $z\sim2$ nebulae. This is consistent with their average SB profiles (e.g., \citealt{Cai2019}).

\begin{table}
\centering
\begin{tabular}{cccccc}
\hline
Data set & $a_{\rm 0}$ & $a_{\rm 1}$ & corr.coeff & $rms$ & SB\\
\hline
\multicolumn{6}{c}{Physical Areas}\\
\hline
all$_{\rm ref}$ & -4.21$\pm$0.13 & 1.13$\pm$0.04 & 0.96 & 0.19 & 1.23\\
inliers$_{\rm ref}$ & -4.23$\pm$0.20 & 1.13$\pm$0.06 & 0.92 & 0.14 & 1.23\\
 \hline
 all$_{\times 2}$ & -4.12$\pm$0.09 & 1.17$\pm$0.03 & 0.98 & 0.14 & 2.46\\s
inliers$_{\times 2}$ &  -4.10$\pm$0.07 & 1.15$\pm$0.03 & 0.98 & 0.11 & 2.46\\
 \hline
all$_{\times 4}$ & -3.93$\pm$0.07 & 1.19$\pm$0.03 &   0.98 & 0.11 & 4.92\\
inliers$_{\times 4}$ & -3.88$\pm$0.07 & 1.17$\pm$0.03 &  0.99 & 0.09 & 4.92\\
 \hline
\multicolumn{6}{c}{Normalized Areas}\\
\hline
all$_{\rm ref}$ & 0.94$\pm$0.06 & 1.09$\pm$0.04 & 0.96 & 0.19 & 1.23\\
inliers$_{\rm ref}$ & 1.04$\pm$0.07 & 1.18$\pm$0.05 & 0.93 & 0.17 & 1.23\\
 \hline
 all$_{\times 2}$ & 1.22$\pm$0.06 & 1.15$\pm$0.03 & 0.97 & 0.15 & 2.46\\s
inliers$_{\times 2}$ &  1.24$\pm$0.06 & 1.16$\pm$0.04 & 0.97 & 0.14 & 2.46\\
 \hline
all$_{\times 4}$ & 1.49$\pm$0.08 & 1.16$\pm$0.04 &   0.97 & 0.15 & 4.92\\
inliers$_{\times 4}$ & 1.45$\pm$0.08 & 1.14$\pm$0.04 &  0.96 & 0.14 & 4.92\\
 \hline
\multicolumn{6}{c}{Comoving Areas}\\
\hline
all$_{\rm ref}$ & -5.38$\pm$0.17 & 1.09$\pm$0.04 & 0.96 & 0.19 & 1.23\\
inliers$_{\rm ref}$ & -5.80$\pm$0.25 & 1.18$\pm$0.05 & 0.93 & 0.17 & 1.23\\
 \hline
 all$_{\times 2}$ & -5.46$\pm$0.14 & 1.15$\pm$0.03 & 0.97 & 0.16 & 2.46\\s
inliers$_{\times 2}$ &  -5.53$\pm$0.18 & 1.16$\pm$0.04 & 0.96 & 0.15 & 2.46\\
 \hline
all$_{\times 4}$ & -5.24$\pm$0.15 & 1.16$\pm$0.04 &   0.97 & 0.15 & 4.92\\
inliers$_{\times 4}$ & -5.14$\pm$0.16 & 1.14$\pm$0.04 &  0.96 & 0.15 & 4.92\\
 \hline
\end{tabular}
\caption{The linear regression fits of Figures~\ref{Fig2}, ${{\rm log_{10}}(L^{\rm Neb}_{\rm Ly\alpha}/[10^{44}\,{\rm erg\ s^{-1}}])= a_1 {\rm log_{10}}({\rm Area}^{\rm Neb})+a_0}$, where ${\rm Area}^{\rm Neb}$ is in ${\rm kpc}^2$, adimensional or ${\rm ckpc}^2$. The normalisations $a_{\rm 0}$ are in units of log(10$^{\rm 44}$erg s$^{\rm -1}$), while, from top to bottom, the slopes $a_{\rm 1}$ are in log(10$^{\rm 44}$erg s$^{\rm -1}$) log(kpc$^{\rm2}$)$^{\rm -1}$, log(10$^{\rm 44}$erg s$^{\rm -1}$), and log(10$^{\rm 44}$erg s$^{\rm -1}$) log(ckpc$^{\rm2}$)$^{\rm -1}$. 
The last column lists the SB threshold for the lowest redshift object considered, in units of $10^{-17}$~erg~s$^{-1}$~cm$^{-2}$~arcsec$^{-2}$.}
\label{Tab1}
\end{table}

\section{Extending the luminosity-area relation to higher redshift quasars}
\label{sec:comparison}

To further confirm that the relation is in place at different redshifts, we separately fit the $z\sim2$ and $z\sim3$ samples and find that the normalizations and slopes remain within 10~\% with respect to the global fit. This result seems to indicate 
that the relation does not change with redshift. We  
test this finding out to the highest redshift quasars.
As stated in Section~\ref{sec:cut} we cannot rely on current observations of Ly$\alpha$ nebulae around $z\sim6-7$ quasars 
as they are too shallow. Therefore, 
we compare the discovered luminosity-area relation with predictions from 
simulations. 
Cosmological, radiation-hydrodynamic simulations  
can capture the relevant processes for a realistic modelling of the Ly$\alpha$ emission in QSO environments, e.g., non-equilibrium ionisation states, photo-ionisation due to QSO and stellar radiation, gravitational accretion shocks, galactic and AGN feedback.
Such simulations are few and far between as they are very computationally intensive (for $z>6$  
\citealt{Costa2022}; or down to $z\sim3$ 
\citealt{Trebitsch2021} but for QSOs at least one order of magnitude less luminous than the ones in our sample).

\citet{Costa2022} post-processed a suite of high-resolution, radiation-hydrodynamic cosmological simulations targeting QSO host haloes with the Ly$\alpha$ radiative transfer code \textsc{Rascas} (\citealt{Michel-Dansac2020}), finding a close match to the average SB profiles and morphologies of observed Ly$\alpha$ nebulae at $z\sim6$. This close agreement between models and observations unambiguously depends on (i) the inclusion of QSO-powered outflows and (ii) modeling of Ly$\alpha$ resonant scattering in post-processing. Taking into account that 
the high-redshift QSOs considered  
here are expected to inhabit similar halos ($M_{\rm DM}\sim 10^{12.5}$~M$_{\odot}$), and that their observed SB profiles are similar in shape 
(once corrected for cosmological dimming and compared in comoving units), we compare the predictions of 
Costa et al. reference simulation
with our observational sample.  This simulation has $M_{\rm DM}=2.4 \times 10^{12}$~M$_\odot$ at $z\sim6$ 
and a QSO with 
$L_{\rm bol}=10^{47.5} \, \rm erg \, s^{-1}$ and a fractional QSO luminosity in the Ly$\alpha$ line of $f_{\rm Ly\alpha}=0.005$. 
We then obtain Ly$\alpha$ maps for 12 galaxy inclinations $i$ ( with respect to our line-of-sight) between $90^{\circ}$ and $0^{\circ}$, equally spaced in cos($i$).  
After we smooth each map to the seeing of the observations, we compute its Ly$\alpha$ luminosity, area, and asymmetry using the SB threshold corresponding to the $z$ of the simulation ($z=6.2$; SB$_{\rm Ly\alpha}=3.9\times10^{-19}$~erg~s$^{-1}$~cm$^{-2}$~arcsec$^{-2}$) and neglecting a circular region of $\sim$1~arcsec$^2$ at the QSO position, as done for the observations. 
We note that this calculation  
considers all powering mechanisms for Ly$\alpha$ including resonant scattering from the QSO's broad-line region (BLR).

\begin{figure}
    \includegraphics[width=1.0\columnwidth]
    {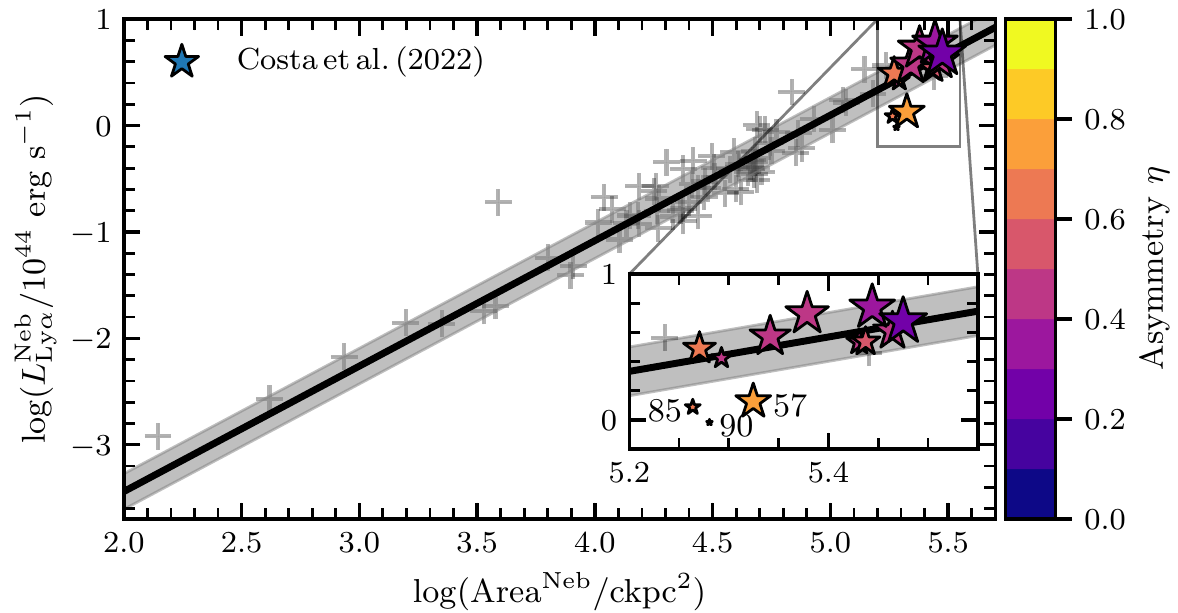}
    \caption{The linear regression  
    for the whole sample with its $rms$ (black line with shaded region) is compared to the 
    simulated data (stars; \citealt{Costa2022}). 
    Each of the simulated data-points, color-coded following their nebula asymmetry, represents a different host-galaxy inclination $i$, 
    and we vary their sizes as a function of ${\rm cos}(i)$. 
    The inset zooms on the simulated data-points, with three of them highlighted by their $i$.} 
    \label{Fig3}
\end{figure}

Figure~\ref{Fig3} shows the results of this exercise in comparison to the linear regression in the case of comoving areas for the whole sample. 
The simulated nebulae end up on the relation, 
and are the largest and brightest.
This is expected from the luminosities of the $z\sim6$ nebulae discovered by \citet{Farina2019}, whose observations are only sensitive to significantly higher SB values than used here (Section~\ref{sec:cut}). The luminosity-area relation seems therefore in place also for the Ly$\alpha$ nebulae around the highest redshift quasars. Importantly, the fainter and smaller simulated nebulae are on average those around the most inclined host-galaxies.

We emphasize that simulations have a limited resolution. The minimum cell size in the most refined halo regions 
is $\simeq80$~pc. Higher, not resolved, densities could therefore increase the Ly$\alpha$ emission  
(e.g., \citealt{cantalupo14}).
However, \citet{Costa2022} showed that scattering from QSO's BLR photons is a viable mechanism and extreme clumping factors may not be necessary to match observations. The detailed balance of the different powering mechanisms requires further observational and theoretical work.

\section{What drives the luminosity-area relation?}
\label{sec:drivers}

The simulated nebulae show a higher degree of asymmetry for large $i$ (i.e. host-galaxy closer to edge-on). According to \citet{Costa2022} such asymmetries arise naturally due to the presence of Ly$\alpha$ escape channels aligned perpendicularly to the galaxy's disk, where the quasar's outflow propagates most efficiently. For large $i$ 
resonant scattering becomes more important, resulting in lopsided, and fainter nebulae. This fact might explain why we observe the more asymmetric nebulae in the lower part of the relation (Figure~\ref{Fig1}).
This prediction could be tested with ALMA by constraining the host-galaxy inclination from resolved molecular gas kinematics.

We stress that the common SB cut used here  
implies that the luminosity-area relation probes different scales at different redshifts. 
In other words, $z\sim2$ Ly$\alpha$ nebulae are cut well within the expected virial radius of the host DM halos, while at $z\sim6$ they extend into the intergalactic medium. It is therefore expected that a $z\sim6$ simulation, as used here, cannot cover the lower part of the relation. 
A large sample of similarly simulated massive halos at different redshifts 
($z>2$) is needed to retrieve the full luminosity-area relation.

Further, we see that 
the two simulated datapoints for $i \gtrsim 85^{\circ}$ 
are clearly not on the observed relation. 
This likely means that an unobscured QSO's host-galaxy cannot be fully edge-on otherwise 
galaxy obscuration would be important. A specific note is required for the simulated data-point for $i=57^{\circ}$ 
as this is showing a similar large deviation from the observed relation as the two points with $i \gtrsim 85^{\circ}$. 
The low Ly$\alpha$ luminosity for this inclination is caused by an absorber belonging to the galaxy disc along this line-of-sight. This 
occurrence 
is not captured by the observed sample which does not include unobscured QSOs with damped Ly$\alpha$ absorbers close to systemic.

Additional processes could account for the very small size of the dimmest observed nebulae. Given the possible importance of the host galaxy and of Ly$\alpha$ resonant scattering
, an obvious complementary explanation could be the presence of more 
dust on galaxy scales in the dimmest systems. \citet{Nahir2022} tentatively show that the dimmest QSO Ly$\alpha$ nebulae are associated with the largest molecular reservoirs, hence with the largest dust masses. Also, it has been argued that "young" QSOs (e.g., \citealt{Eilers2018}) should have smaller extended emission (e.g., \citealt{Farina2019}).
Generalizing, small and dim Ly$\alpha$ nebulae could be due to inefficient or absent quasar feedback (e.g., \citealt{Costa2022}). Observations targeting tracers of AGN outflows (e.g., [OIII]$\lambda$5007, \citealt{Kakkad2020,Vayner2023}) are needed to complement the information from the  
Ly$\alpha$ emission and test this prediction.

\section{Summary}
\label{sec:summary}

We revisit the observations of $z\sim2-3$ QSOs' nebulae (FAB19, \citealt{Cai2019,cantalupo14,hennawi+15})
to study the relation between their luminosity, area and asymmetry by imposing a common dimming-corrected SB threshold.  
We uncover a well-defined luminosity-area relation 
in place at different redshifts ${{\rm log}(L_{\rm Ly\alpha}^{\rm Neb}/[10^{44}\,{\rm erg\ s^{-1}}])=a_1\times{\rm log({\rm Area^{Neb}})}}$ + $a_0$, with most of the asymmetric nebulae located at small luminosities and areas. 

We show that the relation holds for nebulae around the highest redshift quasars
by revisiting cosmological simulations able to reproduce the SB profiles of observed QSOs' nebulae at $z\sim6$ (\citealt{Costa2022}). 
The simulated nebulae 
have luminosities and areas in agreement with the observed relation. 

The comparison with these simulations  
also highlights the possible importance of host-galaxy inclination in shaping the nebulae morphology and brightness, and therefore the relation. 
Type-I QSOs could have their host galaxy at different inclinations, with most 
of them being close to face-on (hence more symmetric and brighter nebulae), while a minority may be more inclined (hence more lopsided and fainter nebulae).
Additional possible processes that could drive the relation and explain small and dim Ly$\alpha$ nebulae are (i) obscuration of the QSO radiation due to the host galaxy, (ii) small age for the QSOs, and (iii) inefficient or absent QSO feedback in some cases.

To fully understand this luminosity-area relation, 
statistical samples of high-resolution cosmological simulations of QSO host galaxies across cosmic time are needed, together with    
larger samples of $z>5$ QSO nebulae observed down to similar $z$-corrected SBs as the sample in this work,  
and observations able to constrain host galaxy properties (e.g., inclination)  
and AGN outflows  
for systems with available Ly$\alpha$ information.

\begin{acknowledgments}
Based on observations collected at the European Organisation for Astronomical Research in the Southern Hemisphere under ESO programmes 094.A-0585(A), 095.A-0615(A), 095.A-0615(B), 096.A-0937(B). Some of the data presented herein 
were obtained at the W. M. Keck Observatory, which is operated as a scientific partnership among the California Institute of Technology, the University of California and the National Aeronautics and Space Administration. The Observatory was made possible by the generous financial support of the W. M. Keck Foundation.
The authors wish to recognize and acknowledge the very significant cultural role and reverence that the summit of Maunakea has always had within the indigenous Hawaiian community.  We are most fortunate to have the opportunity to conduct observations from this mountain.
AO is funded by the Deutsche Forschungsgemeinschaft (DFG, German Research Foundation) –- 443044596.
The work of EPF is supported by NOIRLab, which is managed by the Association of Universities for Research in Astronomy (AURA) under a cooperative agreement with the National Science Foundation.
\end{acknowledgments}

%

\vspace{5mm}
\facilities{VLT(MUSE), Keck(KCWI, LRIS)}


\software{matplotlib \citep{Hunter2007},
astropy \citep{Astropy2018}  
          }





\bibliography{allrefs_LAST}
\bibliographystyle{aasjournal}



\end{document}